\documentclass[12pt,a4paper]{article}
\usepackage{amsmath,amsfonts,amssymb,amscd}
\usepackage[all]{xy}

\makeatletter
\def\@cite#1#2{$^{\mbox{\scriptsize #1\if@tempswa,#2\fi}}$}
\makeatother

\newtheorem{Theorem}{Theorem}
\newtheorem{Definition}[Theorem]{Definition}
\newtheorem{Proposition}[Theorem]{Proposition}

\def\plus#1#2{\vrule height#1pt width0pt depth#2pt}
\def\@{\hskip.8pt}
\def\?{\hskip.3pt}

\def\Ref#1#2{\if#2)\ref{#1}#2\else\ref{#1}\@#2\fi}

\def\eps{\varepsilon}
\def\g{\gamma}

\def\vphi{\varphi}

\def\w{\omega}

\def\P_#1{\mathcal P_{#1}\/}

\def\h#1{\hat{#1}}
\def\j#1{j_1\/(#1)}

\def\nutilde{\?{\scriptstyle\mbox{\tiny\raise-1.6ex\hbox to1pt{$\sim\hss$}}\nu\@\?}}
\def\Nutilde{\raise-1ex\hbox to1pt{$\scriptstyle\sim\hss$} \nu}
\def\phitilde{\raise-1.4ex\hbox to1pt{$\scriptstyle\sim\hss$} \vphi}
\def\Gammatilde{\raise-1ex\hbox to1pt{$\scriptstyle\sim\hss$} \Gamma}
\def\BIG)_#1{\Big)_{\!\@\lower-2.2pt\hbox{$\scriptstyle#1$}}}

\def\Ig0{I\/\big[\?\g_\xi\@\big]_{\lower1.4pt\hbox{$\scriptstyle\xi=0$}}}

\def\Eps#1{\eps_{\lower1pt\hbox{$\scriptstyle #1$}}}

\def\d#1/d#2{\frac{d\/#1}{d\/#2}}
\def\de#1/de#2{\frac{\partial\/#1}{\partial\/#2}}
\def\SD#1/de#2/de#3{\ifx#2 \frac{\plus02\partial^{\@\@2}#1}
    {\plus90\partial\@#3^{\@2}} \else\frac{\plus02\partial^{\@\@2}#1}
    {\partial\?#2\partial\?#3}\fi}
\def\TD#1/de#2/de#3/de#4{\frac{\plus02\partial^{\@\@3}#1}%
    {\partial\?#2\partial\?#3\partial\?#4}}
\def\D#1/D#2{\frac{D\/#1}{D\/#2}}

\def\DD#1/D#2{\textstyle{\text{\Large$\D{#1}/D{#2}$}}}
\def\dd#1/d#2{\textstyle{\text{\Large$\d{#1}/d{#2}$}}}
\def\De#1/de#2{\textstyle{\text{\Large$\de{#1}/de{#2}$}}}
\def\sD#1/de#2/de#3{\textstyle{\text{\Large$\SD{#1}/de{#2}/de{#3}$}}}

\def\iff{\Leftrightarrow}

\def\interior{\,\hbox{\vrule depth0pt height.6pt width4pt%
\vrule depth0pt height8pt}\;\,}

\begin{document}
\vskip-2cm

\title{Constructing a class of solutions for the Hamilton--Jacobi equation in field theory.}
\author{Danilo Bruno \\
        Dipartimento di Matematica dell'Universit\`a di Genova \\
        Via Dodecaneso, 35 - 16146 Genova (Italia) \\
        E-mail: bruno@dima.unige.it
          }
\date{}
\maketitle

\begin{abstract}
\noindent
A new approach leading to the formulation of the Hamilton--Jacobi equation for field theories is investigated within the framework of jet--bundles and multi--symplectic manifolds. An algorithm associating classes of solutions to given sets of boundary conditions of the field equations is provided. The paper also puts into evidence the intrinsic limits of the Hamilton--Jacobi method as an algorithm to determine families of solutions of the field equations, showing how the choice of the boundary data is often limited by compatibility conditions.
\par\bigskip
\noindent
{\bf PACS:} 11.10.Ef\\
{\bf 1991 Mathematical subject classification:} 70H20, 70S05, 35C05
\newline
{\bf Keywords:} Hamiltonian field theories, Hamilton-Jacobi equation. 
\end{abstract}
\thispagestyle{empty}

\section{Introduction}

The Hamilton--Jacobi equation in classical point particle mechanics is a powerful tool, used to solve the evolution equations, by means of the concept of complete integral. Moreover, its theoretical importance is placed in its being the cornerstone of the transition from classical to non--relativistic quantum mechanics in Schroedinger' s formulation. 

An analogue of Hamilton--Jacobi equation for field theory has been created at the beginning of the XIX century through different alternative approaches \cite{Rund}. In particular the formulation of De Donder and the one of Caratheodory have been rather studied and further developed, even recently. Yet none of them has proved to be as powerful as the theory which is generally used in mechanics. In fact, while the equation itself seems to be even less constraining than the mechanical one, being a single PDE in many unknown functions, an additional set of {\it embeddability conditions} is needed to relate its solutions to the field equations. This results into a set of quite restricting conditions and the possibility of finding solutions strongly depends on the nature of the particular problem under consideration. 

It is clear in the literature that the construction of a parametric set of solutions, resembling the classical complete integral, is quite often impossible, due to the presence of strong integrability conditions. Therefore, many different authors pursued the aim of producing sets of particular solutions: some of them built single solutions starting from  given solutions of the field equations \cite{vonRieth,Rovelli}, others use the concept of Baecklund transformation under particular dimensional restrictions \cite{Kastrup}.

The present paper is aimed at building a particular class of solutions of the Hamilton--Jacobi equation, extending the previous approaches. 
What we show here is an algorithm associating particular solutions of the Hamilton--Jacobi equation to given sets of boundary conditions of the field equations. This approach differs from the previous ones in its being independent of any choice of the possible surfaces where the boundary data are assigned and of every dimensional consideration. Moreover there is no need of knowing any particular solution of the field equations in advance. 

Unluckily, the argument is not entirely straightforward, since the above mentioned embeddability conditions play a very strong role and eventually put bounds on the possible initial data which are able to generate solutions of the field equations by way of Hamilton--Jacobi. These topics are examined within the proposed class of solutions, which is by no means the most general one. It is opinion of the author that some generalizations of the algorithm could be possible and could even help weakening some of the restrictions; yet most of the limitations do not actually depend on the particular choices, but on the nature of the problem itself. 

In section 2, the Lagrangian and Hamiltonian aspects of field theories within the framework of jet--bundles are revised\cite{saunders,Sardanashvily,Mangiarotti}. These arguments are an actual survey on currently known arguments and are reported for notational purposes only\cite{cvb4}. 

Section 3 is devoted to the study of the Hamilton--Jacobi equation. After proposing an easy alternative deduction of the equation itself, completely equivalent to De Donder's one, the concept of embeddability of solutions is explained and the conditions for this to happen are worked out. Then, the algorithm leading to the determination of a class of solutions is explained in detail. 

Finally, an easy example is provided in Section 4, with the aim of showing how the method works, avoiding useless computational complications. 

\section{Preliminaries}

This section is devoted to recalling the geometrical environment where Lagrangian and Hamiltonian dynamics for field theories is developed, within the framework provided by jet--bundles. \\[2pt]
(i)\,It is known that field theories are generally described from a geometric viewpoint taking a bundle $\pi: E\to M$ into account, endowed with a set of local coordinates $x^1,\ldots,x^n,y^1,\ldots,y^r$, the first set denoting the basis coordinates on $M$ and the remaining ones the local coordinates on the fibers. A physical field is then represented as a section $\varphi : M\to E$, locally written in the form:
\[
y^i = \varphi^i\/(x^\mu) \quad i=1,\ldots,r \; , \; \mu = 1,\ldots,n \; .
\]
(ii)\,Lagrangian theories are described taking the first jet--bundle $\j E$ associated with $\pi:E\to M$ into account and working on the geometrical framework provided by the double fibration $\j E \stackrel{\h\pi}{\to} E \stackrel{\pi}{\to} M$. For variational purposes, the manifold $M$ is supposed to be endowed with a metric $\phi = g_{\mu\nu} dx^\mu\otimes dx^\nu$. 
\\[2pt]
\noindent
The bundle $\j E$ is endowed with a set of local fibered coordinates $x^\mu,y^i,y^i_\mu$; the last set of coordinates represents the equivalence classes of sections $\varphi$ having a first--order contact. The whole set is subject to the following transformation laws:
\begin{equation}\label{2.1}
\bar x^\mu = \bar x^\mu\/(x^\nu) \; , \; \bar y^i = \bar y^i\/(x^\nu,y^j) \; , \; \bar y^i_\nu = \left(\de \bar y^i /de{x^\mu} + \de \bar y^i /de{y^j}\, y^j_\mu \right) \de x^\mu /de {\bar x^\nu} .
\end{equation}
Eqs.~\eqref{2.1} show that the bundle $\j E\to E$ has the nature of an affine bundle, modelled on $T^*\/(M)\otimes V\/(E)$, where $V\/(E)$ denotes the vector bundle of vertical vector fields on $E\to M$. 

\noindent
The bundle $\j E$ is endowed with a set of canonical differential forms, henceforth called {\it contact} $1$--forms, that result to be invariant under a coordinate transformation as the one described by eq.~\eqref{2.1}. They can be locally described in the form:
\begin{equation}\label{2.2}
\w^i = dy^i - y^i_\mu \, d x^\mu \quad .
\end{equation}

\noindent
Every section $\varphi:M\to E$ can be raised to a section $\j\varphi : M \to \j E$ as 
\[
y^i = \varphi^i\/(x^\mu) \quad , \quad y^i_\mu = \de \varphi^i /de {x^\mu} \/(x^\mu) \quad .
\]
Conversely, every section $\h\varphi:M\to \j E$ is said to be {\it admissible} if there exists a section $\varphi : M\to E$ such that $\h\varphi = \j\varphi$. It is easy to prove that a section is admissible if and only if $\h\varphi^*\/(\w^i) = 0$ \quad . \\[2pt]
(iii)\,Lagrangian field theories are described assigning a function $L\/(x^\mu,y^i,y^i_\mu)$ on $\j E$, called the {\it Lagrangian}, accounting for the dynamical aspects. Depending on the theory under consideration, the choice of the Lagrangian is generally made following geometrical guidelines, symmetry properties or a constitutive analysis of the interactions. In this work we will sistematically suppose this choice has been made, and we will be only interested in its mathematical consequences. 

\noindent
Equally important is the presence of a canonical $n$--form $\theta_{L}$ on $\j E$, called the Poincar\`e--Cartan form, induced by the presence of the Lagrangian function and locally represented as
\begin{equation}\label{2.4}
\theta_L = L ds + \de L /de{y^i_\mu} \omega^i \wedge d s_\mu\quad , 
\end{equation}
where $ds := \sqrt{g}\, dx^i\wedge \ldots \wedge dx^n$ is the volume form on $M$ and $ds_\mu := \de /de{x^\mu} \interior ds$. 

\noindent
The latter is invariant under arbitrary coordinate transformations \eqref{2.1}. Actually, its importance lies in the fact that the variational principle for the functional $I\/[\h\varphi] = \int \h\varphi^* (\theta_L) $ on $\j E$ singles out all the admissible sections  $\varphi:M\to E$ satisfying the Euler--Lagrange equations
\begin{equation}\label{2.3}
\de /de{x^\mu} \left[\j{\varphi}^*\left(\de L /de {y^i_\mu} \right)\right] - \de L /de{y^i}\bigg|_{\j\varphi} = 0 \quad . 
\end{equation}
They have the nature of a set of second order partial differential equations for the functions $y^i\/(x^\mu)$, denoting the local representatives of the section $\varphi$. 
\\[2pt]
\noindent
(iv)\,Hamiltonian field theory is developed within the framework provided by the bundle $\Lambda^n\/(E)$ of skew--simmetric $n$-forms on $E$ (compare with \cite{cvb4}). Let us take the bundle $\Lambda^n_1\/(E)$ of horizontal $n$-forms into account, whose elements are annihilated whenever one of its arguments is vertical. Then, let $\Lambda^n_2\/(E)$ be the bundle of $1$-contact forms on $E$, whose elements have the property of vanishing when at least two of their arguments are vertical vector fields. 
The elements of the two bundles can be locally represented by the following conditions:
\begin{subequations}\label{2.6}
\begin{equation}
\sigma \in \Lambda^n_1\/(E) \quad \iff \quad \sigma = p_0\/(\sigma)\,ds \quad , 
\end{equation}
\begin{equation}
\sigma \in \Lambda^n_2\/(E) \quad \iff \quad \sigma = p\/(\sigma)\,ds + p^\mu_i\/(\sigma)\,dy^i\wedge ds_\mu \quad . 
\end{equation}
\end{subequations}
The bundles $\Lambda^n_1\/(E)$ and $\Lambda^n_2\/(E)$ are respectively described by a system of local coordinates $x^\mu,y^i,p_0$ and $x^\mu,y^i,p,p^\mu_i$, subject to the following transformation laws:
\begin{subequations}\label{2.7}
\begin{equation}
p_0 = J\, \bar p_0 \quad , \quad {\rm where}\; J := \det\left\|\de{(x^1,\ldots,x^n)}/de{(\bar x^1,\ldots,\bar x^n)}\right\|
\end{equation}
\begin{equation}
p = J\left(\bar p + \bar p^\rho_j \de\bar y^j /de{x^\lambda}\de x^\lambda/de{\bar x^\rho} \right) \quad ; \quad p^\mu_i = J\, \de \bar y^j /de{y^i} \de x^\mu /de{\bar x^\rho} \bar p^\rho_j
\end{equation}
\end{subequations}
The latter shows that $\Lambda^n_1\/(E)\to E$ is a vector sub--bundle of  $\Lambda^n_2\/(E) \to E$;  the quotient bundle $\Pi\/(E):=\Lambda^n_2\/(E) / \Lambda^n_1\/(E)$ will be henceforth called the {\it phase space} and endowed with a set of local coordinates $x^\mu,y^i,p^\mu_i$. The transformation laws (\ref{2.7}b) make $\Lambda^n_2\/(E)\to \Pi\/(E)$ into an affine bundle, modelled on $\Lambda^n_1\/(E)$. \\[2pt]
(v)\,The bundle $\Lambda^n_2\/(E)$ is endowed with a canonical Liouville $n$--form $\Theta$, locally expressed as
\begin{equation}\label{2.8}
\Theta = p\,ds + p^\mu_i\,dy^i\wedge ds_\mu \quad , 
\end{equation}
whose differential is represented by
\begin{equation}\label{2.9}
\Omega = d\Theta = dp\wedge ds + dp^\mu_i \wedge dy^i \wedge ds_\mu\quad . 
\end{equation}
The latter is a multi--simplectic $(n+1)$--form on $\Lambda^n_2\/(E)$ endowing it with a non--singular multi--symplectic structure. \\[2pt]
(vi)\,The presence of the Liouville $n$--form allows to write the field equations \eqref{2.3} on $\Lambda^n_2\/(E)$. In fact, there exists a unique fibered differentiable application $\lambda: \j E \to \Lambda^n_2\/(E)$ such that $\lambda^*\/(\Theta) = \theta_L$. 
This requirement can be translated into local fibered coordinates, representing the map $\lambda$ in the form:
\begin{equation}\label{2.10}
\lambda:\quad  p^\mu_i = \de L/de{y^i_\mu} \quad , \quad p = L\/(x^\mu,y^i,y^i_\mu) -  \de L/de{y^i_\mu}  y^i_\mu \quad . 
\end{equation}
The application $\lambda:\j E\to \Lambda^n_2\/(E)$ represents a sub--manifold of $\Lambda^n_2\/(E)$, fibered on the phase space $\Pi\/(E)$;
 its image will be denoted by ${\cal S} := \lambda\/(\j E)$. In particular, whenever the regularity condition 
\begin{equation}\label{2.11}
\det\left|\SD L/de y^\mu_i /de{ y^\nu_j} \right| \neq 0 
\end{equation}
is satisfied, the first equation \eqref{2.10} can be locally inverted, allowing to obtain the coordinates $y^i_\mu = y^i_\mu\/(x^\nu,y^i,p^\mu_i)$. If the latter is substituted into the second equation \eqref{2.10}, the following cartesian representation for the manifold ${\cal S}$ is obtained:
\begin{equation}\label{2.12}
{\cal S} : \quad p = L\/(x^\mu,y^i,y^i_\mu\/(x^\nu,y^i,p^\mu_i)) - y^i_\mu\/(x^\nu,y^i,p^\mu_i) p^\mu_i = - H\/(x^\mu,y^i,p^\mu_i) \quad . 
\end{equation}
The function $H\/(x^\mu,y^i,p^\mu_i)$ will be called the Hamiltonian of the system: it actually is (the pull--back of) a function on $\Pi\/(E)$. 

\noindent
A comparison with eq.~\eqref{2.12} shows that whenever the regularity condition \eqref{2.11} is fulfilled, the sub--manifold ${\cal S}$ has the nature of a local section of the affine bundle $\Lambda^n_2\/(E)\to \Pi\/(E)$, locally represented as:
\begin{equation}\label{2.13}
h : \Pi\/(E) \to \Lambda^n_2\/(E) \quad ; \quad h: p + H\/(x^\mu,y^i,p^\mu_i) = 0 \quad . 
\end{equation}
The latter will be henceforth called the {\it Hamiltonian section} and its assignment is completely equivalent to the knowledge of the Lagrangian function on $\j E$, because of eqs.~\eqref{2.10} and \eqref{2.12}. \\[2pt]
(vii)\,Finally, the presence of the Hamiltonian section allows to pull--back the Liouville $1$--form on $\Pi\/(E)$ and to build a variational principle on the phase space. Consider a section $\tilde\varphi: M \to \Pi\/(E)$, locally represented in the form
\begin{equation}\label{2.14}
\tilde\varphi : \quad y^i = y^i\/(x^\mu) \quad , \quad p^\mu_i = p^\mu_i\/(x^\mu) \quad . 
\end{equation}
An easy calculation shows that the stationary sections for the functional
\[
I_h\/[\tilde\varphi] = \int \tilde\varphi^*h^*\/(\Theta)
\]
are the solutions of the following differential equations
\begin{equation}\label{2.15}
\de y^i /de{x^\mu} = \de H /de {p^\mu_i} \quad ; \quad \de p^\mu_i/de{x^\mu} = - \de H/de{y^i} \quad , 
\end{equation}
which are the equivalent of Hamilton equations in field theories. They have the nature of a system of first order PDEs in the unknowns $y^i$ and $p^\mu_i$ and are often harder to solve than their Lagrangian counterpart, because of the difficulties needed to uncouple them. 

\noindent
Every solution $\tilde\varphi$ of the sistem \eqref{2.15} is $\lambda$--related to the solutions of eqs.~\eqref{2.3}:
\[
\lambda\cdot \j\varphi = \tilde \varphi \quad . 
\]
This means that the determination of a solution $\tilde\varphi$ of \eqref{2.15} is equivalent to determining a solution of eq.~\eqref{2.3}. Every solution $\tilde \varphi:M \to \Pi\/(E)$ of eqs.~\eqref{2.15} will be called a {\it critical field}. The determination of an alternative method yielding solutions of \eqref{2.15} is the principal aim of the present paper. 

\section{The Hamilton-Jacobi equation}

The usual deduction of the Hamilton--Jacobi equation in analytical mechanics is based on the use of canonical transformations on symplectic manifolds by means of the concept of generating function. Unluckily, no such theory is available in the multisymplectic case. Our deduction of the Hamilton--Jacobi equation will be therefore based on an alternative approach, relying on the possibility of determining a particular sub--manifold of the surface ${\cal S}$, containing the lift of a solution of the field equations with the given boundary conditions. This requirement singles out a first order PDE and a set of additional {\it embeddability conditions}. 

\noindent
Let us consider the geometric structures determined by the following construction. Let $\sigma\in\Lambda^{n-1}_{1}\/(E)$ be a generic horizontal $(n-1)$-form on $E$, locally described in the form:
\begin{equation}\label{3.1}
\sigma = S^\alpha\/(x^\mu,y^i)\,ds_\alpha\quad . 
\end{equation}
Its exterior differential $d\sigma\in\Lambda^n_2\/(E)$ is locally described as:
\begin{equation}\label{3.2}
d\sigma = \de S^\alpha\/(x^\mu,y^i)/de {x^\alpha} ds + \de S^\alpha\/(x^\mu,y^i)/de{y^i} dy^i\wedge ds_\alpha \quad . 
\end{equation}
It represents a section $\Sigma: E\to \Lambda^n_2\/(E)$ (compare with (\ref{2.6}b)) locally written as:
\begin{equation}\label{3.3}
\Sigma: \quad p =  \de S^\alpha\/(x^\mu,y^i)/de {x^\alpha} \quad , \quad p^\alpha_i =  \de S^\alpha\/(x^\mu,y^i)/de{y^i}  \quad . 
\end{equation}
Such a section is the natural generalization of the concept of Lagrangian sub--manifold to the multi--symplectic case, since the pull--back of the Liouville $n$--form to $\Sigma$ is exact:
\[
\Sigma^*\/(\Theta) = \de S^\alpha/de {x^\alpha} ds + \de S^\alpha/de{y^i} dy^i \wedge ds_\alpha = dS^\alpha\/(x^\mu,y^i) \wedge ds_\alpha = d\/(\sigma) \quad . 
\]
We are now going to determine under what circumstances the above section $\Sigma:E\to \Lambda^n_2\/(E)$ is contained in the sub--manifold ${\cal S}\subset\Lambda^n_2\/(E)$.

\noindent
A comparison between eq.~\eqref{2.12}, defining the sub--manifold ${\cal S}$, and eq.~\eqref{3.3} shows that the following condition must necessarily be satisfied:
\begin{equation}\label{3.4}
\de S^\alpha/de{x^\alpha} + H\/\left(x^\mu,y^i,\de S^\alpha/de{y^i}\right) = 0 \quad . 
\end{equation}
The latter is known in the literature as {\it Hamilton--Jacobi equation} for field theory. This equation is deeply linked with the solutions of the field equations; in fact, 
performing a derivative of both sides of \eqref{3.4} with respect to $y^i$, we obtain that: 
\[
\de /de{x^\alpha} \de S^\alpha /de{y^i} + \de H /de{y^i} + \SD S^\alpha /de y^i /de {y^j} \de H /de {p^\alpha_j} = 0 \;  \Rightarrow \; \de /de{x^\alpha} \left( \de S^\alpha /de{y^i}\/(x^\mu,y^i\/(x^\mu)) \right)  = - \de H /de{y^i} \quad . 
\]
This means that the quantities $p^\alpha_i = \de S^\alpha /de{y^i}$ fulfill the second set of eqs.\eqref{2.15} for every $y^i = y^i\/(x^\mu)$. Substituting them into the first set of \eqref{2.15} we obtain the condition:
\begin{equation}\label{3.4b}
\de y^i /de{x^\mu} = \de H /de{p^\mu_i} \left(x^\mu, y^i,\de S^\mu /de {y^i} \right) \quad . 
\end{equation}

\noindent
Eq.~\eqref{3.4b} is not identically satisfied by all the solutions of the Hamilton--Jacobi equation, but is needed in order to make the field equations valid. 

\begin{Definition}
A solution $y^i = y^i\/(x^\nu)$, $p_i^\mu = p_i^\mu\/(x^\nu)$ is said to be embeddable in a solution $S^\mu = S^\mu\/(x^\nu,y^i)$ of \eqref{3.4} if condition \eqref{3.4b} holds. 
\end{Definition}

\noindent
If $\mu=1$, i.e. in the case of point particle mechanics, eq.~\eqref{3.4b} describes a velocity field on $E$ for every solution of \eqref{3.4} and its integrability is guaranteed by Cauchy theorem; therefore, it is possible to determine a family of solutions of the field equations embedded in every solution of the Hamilton--Jacobi equation, at least locally.  In the general case, instead, different scenarios are possible.\\[2pt]
$1.$ If the integrability conditions for eq.~\eqref{3.4b} are satisfied, then a family of  solutions of the field equations can be embedded in a solution $S^\mu\/(x^\mu,y^i)$. This condition is called in the literature {\it strong embeddability condition} and is generally satisfied in much limited situations only; however this condition is absolutely required to build a complete analogue with the point--particle mechanics, i.e. to determine a complete integral of the Hamilton--Jacobi theory. This aspects of the theory will not be analyzed here.\\[2pt]
$2.$ In general, the integrability conditions are not satisfied and this gives rise to a restriction on the set of possible solutions. In particular, the generic solution of the Hamilton--Jacobi equation is not able to single out a parametric set of solutions matching all the possible boundary data. This limitation is intrinsic with the nature of the equation itself and an analysis of the possible solutions needs to be handled case by case.\\[2pt]
$3.$ If a generic solution of the Hamilton--Jacobi equation is given, condition \eqref{3.4b} can be checked on it\cite{vonRieth}; this last possibility is called {\it weak embeddability condition} and ensures the validity of the field equations only on the accounted solution $y^i = y^i\/(x^\mu)$. The problem with this approach is that if the integrability conditions for eq.~\eqref{3.4b} are not satisfied around the given critical field, there is no way of using the obtained solution of the Hamilton--Jacobi equation to generate other critical fields. Therefore, this approach has no algorithmic application, since solving the Hamilton--Jacobi equation becomes useless if its only embedded solution is already known. \\[2pt]

\noindent
The approach adopted here will follow an intermediate viewpoint: we will determine a parametric set of solutions of the Hamilton--Jacobi equation, associated to a set of given boundary conditions. This is made possible by the fact that  eq.~\eqref{3.4} has the nature of a single first--order partial differential equation in the $n$ unknown functions $S^\alpha\,(x^\mu,y^i)$ and therefore possesses a great amount of equivalent solutions. This arbitrariness will be eventually fixed by the weak embeddability condition. Since only the solution associated with the given boundary data is sought, no integrability condition for \eqref{3.4b} is required. 

\noindent
In order to associate a solution to a set of boundary conditions, we first remind that 
they  essentially consist in the assignment of the fields and of their normal derivative on an $(n-1)$-dimensional submanifold $i:N\to M$, locally expressed as 
\[
i: x^\mu = \varphi^\mu\/(\zeta^1,\ldots,\zeta^{n-1}) \quad , \quad \mu = 1\ldots n \quad , 
\]
where $\zeta^1,\ldots,\zeta^{n-1}$ represent a system of local coordinates on $N$. The whole set of boundary data is therefore represented as:
\begin{equation}\label{3.5}
\left\{
\begin{aligned}
&x^\mu = \varphi^\mu\/(\zeta^A)\quad &&  A=1\ldots n-1 \\
&y^i = \psi^i\/(\zeta^A)\quad  &&  \mu = 1\ldots n \\
&\de y^i/de{x^\mu} n^\mu = \h\psi^i\/(\zeta^A)\quad  && i=1\ldots r
\end{aligned}
\right.\quad , 
\end{equation}
where $n = n^\mu \de /de{x^\mu}$ is a vector field defined on $N$ and transverse to it. 

\noindent
Now, let us consider the particular solution of the Hamilton--Jacobi equation obtained through the following procedure.  
Let  $X = X^\mu\/(x^\nu) \de /de{x^\mu} \in D^1\/(M)$ be an arbitrary vector field having the property of being transverse to the surface of the boundary data (but not necessarily equal to $n$ on $N$) and take the following class of candidate solutions of the Hamilton--Jacobi equation into account
\begin{equation}\label{3.6}
S^\mu\/(x^\nu,y^i) := \varphi(x^\nu,y^i)\,X^\mu(x^\nu) \quad , 
\end{equation}
where $ \varphi(x^\mu,y^i)$ is an arbitrary function on $E$. 
Eq.~\eqref{3.6} fullfills the Hamilton--Jacobi equation if the function $\varphi\/(x^\mu,y^i)$ is such that
\begin{equation}\label{3.7}
\de \varphi /de {x^\mu} X^\mu + \varphi \de X^\mu /de{x^\mu} + H\/(x^\mu, y^i, \de \varphi /de {y^i} X^\mu ) = 0 \quad . 
\end{equation}
Eq.~\eqref{3.7} leads to a great simplification of the problem; in fact, it has the nature of a single first order partial differential equation in the single unknown $\varphi(x^\mu,y^i)$ and can be easily solved using the same techniques as the traditional Hamilton--Jacobi equation (e.g. the separation of variables), for every choice of the vector field $X$. 

\noindent
We will now show how determining the solutions of eq.~\eqref{3.7} can be useful to evaluate the critical fields. For this purpose, we will develop a solution algorithm, following the steps below:\\[2pt]
$\bullet$ we think of the components of the vector field $X^\mu$  as free parameters and determine a set of solutions $\varphi = \varphi(x^\mu,y_i,X^\mu)$, depending on arbitrary integration constants;\\[2pt]
$\bullet$ we search for (at least) a vector field $X^\mu = X^\mu(x^\nu)$ satisfying the weak embeddability condition on the given solution; this procedure is partly dependent on the choice of the boundary conditions, since the vector field $X$ needs to be transverse to the surface $N$;\\[2pt]
$\bullet$ we calculate the integration constants, imposing that the solution matches the boundary conditions.\\[2pt]
The parametric role played by the quantities $X^\mu$ suggests that they should appear {\it algebrically} in eq.~\eqref{3.7}. Therefore we choose to impose the condition 
\begin{equation}\label{3.7b}
\de X^\mu /de{x^\mu} = 0 
\end{equation}
on the vector field $X$; it makes the weak embeddability conditions harder, but simplifies the analysis and the solution eq.~\eqref{3.7}.

\noindent
We will cope with the Hamilton--Jacobi equation using the method of characteristics for first--order partial differential equations, stating the argument from a geometric viewpoint\cite{Choquet}.  Let us take the manifold $T\/(E)$ into account: the latter is endowed with a set of local coordinates $x^\mu,y^i,X^\mu$ and these are the actual variables appearing in eq.~\eqref{3.7}. In fact, the choice provided by eq.~\eqref{3.7b} avoids the necessity of using an additional set of variables $X^\mu_\nu$.

\noindent
Let $P\to T\/(E)$ denote a bundle of scalar functions over $T(E)$ and let $u$ be the coordinate along the fiber. Every solution of the differential equation can be thought as a section $\varphi: T\/(E) \to P$. Once the choice of the base manifold $T(E)$ has been made, the presence of eq.~\eqref{3.7b} also avoids the necessity of considering the quantities $X^\mu$ among the unknowns of the differential equation: this would actually bring us back to the original problem of having multiple unknowns. 

\noindent
Provided the above choices are made, the first jet--bundle $\j P$ is then endowed with a set of local fibered coordinates $x^\mu, y^i,X^\mu,u,u_\mu,u_i,U_\mu$, with the formal identifications $u_\mu \simeq \de u /de {x^\mu} ,u_i \simeq \de u /de{y_i} ,U_\mu \simeq \de u /de{X^\mu} $. 
This is a classical context where PDEs can be framed from a geometrical viewpoint, interpreting the differential equation as a sub-manifold of $\j P$, locally written in cartesian form as 
\begin{equation}\label{3.8}
F(x^\mu, y^i,X^\mu,u,u_\mu,u_i,U^\mu) = u_\mu X^\mu +  H\/(x^\mu,y^i,u_i\,X^\mu) = 0 \quad . 
\end{equation}
We notice that eq.~\eqref{3.8} independent of the variables $U_\mu$. The characteristic curves can be calculated solving the usual system of first order ordinary differential equations on $\j P$:
\begin{equation}\label{3.9}
\begin{aligned}
&\d x^\mu/d\xi = \de F/de{u_\mu} \quad ; \quad \d y^i /d\xi = \de F/de{u_i}  \quad ; \quad \d X^\mu/d\xi = \de F/de{U_\mu} ;\\[2pt]
&\d u /d{\xi} = \de F/de u + u_\mu \de F/de{u_\mu} + u_i \de F/de{u_i} +  U_\mu \de F/de{U_\mu} \\[2pt]
&\d u_\mu/d\xi = - \de F/de{x^\mu} - \de F/de u u_\mu  \quad;\quad \d u_i/d\xi = - \de F/de{y^i}  - \de F/de u u_i \quad ; \quad \d U_\mu/d\xi = - \de F/de{X^\mu} - \de F/de u U_\mu 
\end{aligned}
\end{equation}
In the present case, taking eqs.~\eqref{3.6} and \eqref{3.7} into account, the above equations can be evaluated as follows:
\begin{subequations}\label{3.10}
\begin{equation}
\d X^\mu /d{\xi} = 0 
\quad ; \quad 
\d x^\mu /d\xi = X^\mu
\end{equation}
\begin{equation}
\d y^i/d\xi = X^\mu\de H/de{p^\mu_i}\/(x^\mu,y^i,u_i\,X^\mu) 
\quad ; \quad 
\d u_i /d\xi = - \de H/de {y^i}\/(x^\mu,y^i,u_i\,X^\mu) 
\end{equation}
\begin{equation}
\d u_\mu/d\xi =  - \de H/de{x^\mu} 
\quad ; \quad 
\d u /d \xi = u_\mu X^\mu + u_i \de H/de{p^\mu_i} X^\mu 
\end{equation}
\end{subequations}

\noindent
The first equations (\ref{3.10}a) allows to determine the vector fields $X^\mu$, which result to be constant along the characteristics, and show that the projection on $M$ of the characteristic curves are represented by their integral lines.
\begin{equation}\label{3.11}
X^\mu = A^\mu\/(\zeta^B) \quad \Rightarrow \quad x^\mu = A^\mu\/(\zeta^A) \,\xi + x^\mu_0\/(\zeta^A) 
\end{equation}
The quantities $x^\mu_0\/(\zeta^A)$ can be easily evaluated by imposing $\xi=0$ in \eqref{3.11} and comparing with the boundary conditions \eqref{3.5}. Moreover, it is also possilble to locally
determine the values of $\xi$ and $\zeta^A$ as functions of $x^\mu$ whenever the condition
\begin{equation}\label{3.12}
\det \left(
\begin{array}{c}
A^\mu\/(\zeta^A)  \\
\De A^\mu /de{\zeta^B} \xi + \De x^\mu_0 /de {\zeta^B} 
\end{array}
\right) \neq 0 
\end{equation}
is satisfied. Being the quantities $\zeta^A$ independent coordinates on $N$, it is easy to prove that eq.~\eqref{3.12} is satisfied whenever the vector field $X$ is transverse to the sub--manifold $N$ itself. 

\noindent
Substituting eq.~\eqref{3.12} into (\ref{3.10}b) we obtain 
a system of $2n$ ordinary differential equations of the first order in the unknowns $y^i\/(\xi,\zeta^A,X^\mu),u_i\/(\xi,\zeta^A,X^\mu)$, whose local solvability is guaranteed by Cauchy theorem. Notice that the variables $\zeta^A,X^\mu$ only play a parametric role, being the latter ones independent of $\xi$. The solution is provided by the families of functions 
\begin{equation}\label{3.13}
\begin{aligned}
y^i =& f^i\/(\xi,\zeta^A,X^\mu,\alpha_i,\beta^i) \\
u_i =& g_i\/(\xi,\zeta^A,X^\mu,\alpha_i,\beta^i)
\end{aligned}
\end{equation}
written in terms on $2n$ integration constants $\alpha_i,\beta^i$. We now possess all the elements that are needed to impose the embeddability conditions \eqref{3.4b}. Taking the regularity condition \eqref{3.12} into account, they are equivalent to imposing
\begin{subequations}\label{3.14}
\begin{equation}
\de y^i /de{x^\mu} \de x^\mu /de \xi = \de H /de {p^\mu_i} \de x^\mu /de {\xi} \Rightarrow 
\d f^i /d{\xi}  = \de H /de {p^\mu_i} X^\mu\quad ,
\end{equation}
which is identically satisfied because of (\ref{3.10}b) and
\begin{equation}
\de y^i /de{x^\mu} \de x^\mu /de {\zeta^A} = \de H /de {p^\mu_i} \de x^\mu /de {\zeta^A}
\Rightarrow \de f^i /de{\zeta^A}  = \de H /de {p^\mu_i}\/(x^\mu,f^i,g_i\,X^\mu) \de x^\mu /de {\zeta^A}\quad , 
\end{equation}
\end{subequations}
that needs to be imposed by hand on the solution \eqref{3.13}. Notice that the latter is already dependent on the choice of the boundary data. 

\noindent
As a matter of fact, eqs.~(\ref{3.14}b) are a set of $r\times(n-1)$ equations and the existence of a vector field $X^\mu = X^\mu\/(\zeta^A)$ fulfilling them is not guaranteed. If this is not the case, a solution may still be determined adding some of the integration constants $\alpha_i,\beta^i$ among the unknowns. We remark that this drawback is intrinsic in the nature of the problem, since condition \eqref{3.4b} does not depend on the particular algorithm we use, but on the possibility of embedding the critical fields into the solution of the Hamilton-Jacobi equation. On the other hand, the translation of eq.~\eqref{3.4b} into (\ref{3.14}b) depends on the particular choice represented by eq.~\eqref{3.7}: the existence of other possible algorithms leading to a weaker embeddability condition cannot be excluded.   

\noindent
In the follow up, we will suppose that a vector field $X^\mu$ satisfying (\ref{3.14}b) exists, though not excluding that its existence  imposes an eventual restriction on the integration constants. Then,  the following proposition holds:
\begin{Proposition}
Let $X^\mu = X^\mu\/(x^\nu)$ be a vector field satisfying eq.~\eqref{3.7b}. For every choice of $X^\mu$ satisfying eq.~(\ref{3.14}b), and for every $\alpha_i,\beta^i$ (eventually restricting to the ones making eq.~(\ref{3.12}b) solvable), the functions
\[
y^i = f^i\/(\xi(x^\mu),\zeta^A\/(x^\mu),X^\mu(\zeta^A\/(x^\nu)),\alpha_i,\beta^i)
\]
are a solution of the field equations. 
\end{Proposition}

\noindent
The remaining problem is that of matching the above solution with the given boundary conditions \eqref{3.5}. If no restriction on the possible choices of the integration constants $\alpha_i,\beta^i$ is present, they can be determined for every choice of the boundary data on the given surface. Otherwise, eq.~(\ref{3.14}b) singles out a compatibility condition on the boundary data for each given solution and in some cases the problem could even turn out to be impossible. Up to now, this situation has still to be handled case by case. 

\noindent
The solution of the Hamilton--Jacobi can be obtained integrating eqs.~(\ref{3.10}c): the first equation (\ref{3.10}c) allows to determine the quantities $u_\mu = u_\mu(\xi,\zeta^A,\gamma_\mu)$, associated with the given solution and the given boundary conditions, the quantities $\gamma_\mu$ being arbitrary integration constants. 

\noindent
Substituting into the second equation (\ref{3.10}c) we can obtain the solution of the Hamilton--Jacobi equation evaluating the integral:
\begin{equation}\label{3.15}
u\/(\xi,\zeta) = \int \left( u_\mu X^\mu + u_i \de H/de{p^\mu_i} X^\mu \right)\, d\xi \quad . 
\end{equation}
Taking eqs.~\eqref{3.5} and \eqref{3.11} into account, we can write the solution as
\begin{equation}\label{3.16}
S^\mu(x^\nu) = u\/(\xi\/(x^\nu),\zeta^A\/(x^\nu))\, X^\mu\/(\zeta^A\/(x^\nu)) \quad . 
\end{equation}

\noindent
We remark that eqs.~\eqref{3.10} are not linear in the quantities $X^\mu$; therefore no superposition is possible among different solutions obtained using different fields satisfying the embeddability conditions. \\

\noindent
This paper suggests a method to find a certain class of solutions of the Hamilton--Jacobi equation, extending the ones that are currently available in the literature; as far as the present status of the topic is concerned, this does not represent the most general  approach possible. Further developments could be eventually achieved modifying the form of the given solution \eqref{3.6} in its dependence in the $\varphi$ and $X^\mu$. Unluckily, every tentative beyond the linear dependence in the $X^\mu$ complicates the relationship between the characteristics and the integral lines of the vector field $X$: the problem still needs to be analyzed in detail. 

\section{An example}

\noindent
We present here the simplest example possible, in order to show how the method works and what kind of solutions can obtained, avoiding unnecessary techical complications. For this reason, the example is limited to $dim\/(M) = 2$ and the metric is taken to be flat, diagonal and positive definite. None of the above simplifications is needed to make the algorithm work. 

\noindent
Consider a free scalar field $y = y\/(x^1,x^2)$, whose Hamiltonian can be written as
\[
H\/(x^1,x^2,y,p^1,p^2) = \frac{1}{2} \left(\delta_{\mu\nu} p^\mu p^\nu + \mu^2 y^2 \right) \quad , \quad \mu=1,2\quad . 
\]
The Hamilton--Jacobi equation takes the form
\[
\de S^\mu /de {x^\mu} + \frac{1}{2} \left(\delta_{\mu\nu} \de S^\mu /de y\de S^\nu /de y + \mu^2 y^2 \right) = 0 \quad . 
\]
Taking eq.~\eqref{3.6} into account, the latter can be written as
\[
\de \varphi /de{x^\mu} X^\mu + \frac{1}{2} \delta_{\mu\nu} X^\mu X^\nu \left(\de \varphi /de y \right)^2 + \frac{1}{2} \mu^2 y^2 = 0 \quad . 
\]
This last equation could be solved through the separation of variables. Instead, we will follow the guidelines provided in the paper. 

\noindent
Let us suppose the boundary data are given on the surface $ x^1\/(z) = 0, \; x^2\/(z) = z $ in the form:
\begin{equation}\label{4.1}
y(0,z) = f(z) \quad ; \quad \de y /de{x^1}\/(0,z) = g(z)  \quad . 
\end{equation}
Taking the  boundary conditions as well as eqs.~\eqref{3.11} into account we have that the projection on $M$ of the characteristic curves can be locally expressed as functions of the components $A^1$ and $A^2$ of the vector field $X$ as:
\begin{equation}\label{4.2}
x^1\/(z,\xi) = A^1\/(z)\,\xi \quad ; \quad x^2\/(z,\xi) = A^2\/(z)\,\xi + z \quad . 
\end{equation}
This last equation can be used to locally invert the Jacobian matrix $\big|\de (x^1,x^2) /de {(\xi,z)}\big|$ and to calculate the partial derivatives with respect to $\xi,z$. This allows to impose condition \eqref{3.7b} on the field $X$ as follows:
\begin{equation}\label{4.3}
A^1\/(z) \d A^2(z) /d z = A^2\/(z) \d A^1(z) /d z \quad \Rightarrow \quad A^2\/(z) = e^c \, A^1\/(z) \quad . 
\end{equation}
We can now write eqs.~(\ref{3.10}b) explicitly:
\[
\d y(\xi) /d \xi = (\delta_{\mu\nu} X^\mu X^\nu)\, u_1(\xi) \quad ; \quad \d u_1(\xi) /d \xi = - m\, y(\xi) \quad . 
\]
They give rise to the following sets of solutions:
\begin{subequations}\label{4.4}
\begin{equation}
y(\xi,z) = a(z) \cos(\mu\,\alpha(z)\,\xi) + b(z) \sin(\mu\,\alpha(z)\,\xi) 
\end{equation}
\begin{equation}
u_1(\xi,z) = \frac{\mu}{\alpha(z)} \left(b(z) \cos(\mu\,\alpha(z)\,\xi) - a(z) \sin(\mu\,\alpha(z)\,\xi) \right)
\end{equation}
\end{subequations}
where 
$$
\alpha(z) = \sqrt{\delta_{\mu\nu}X^\mu X^\nu} = \sqrt{(A^1\/(z))^2+(A^2\/(z))^2} = A^1\/(z) \sqrt{1+e^{2c}} \quad . 
$$
The weak embeddability condition gives rise to the following system:
\begin{subequations}\label{4.5}
\begin{equation}
\alpha(z) \d \alpha(z) /d z = A^1\/(z) \d A^1\/(z) /d z (1+e^{2c}) \quad , 
\end{equation}
\begin{equation}
\d a(z) /d z = e^c A^1\/(z) \frac{\mu}{\alpha(z)} b(z)\quad , 
\end{equation}
\begin{equation}
\d b(z) /d z = - e^c A^1\/(z) \frac{\mu}{\alpha(z)} a(z) \quad . 
\end{equation}
\end{subequations}
Eq.~(\ref{4.5}a) puts a relationship between $\alpha(z)$ and $A^1\/(z)$ and can be easily integrated as follows:
\[
A^1\/(z) = \frac{\alpha(z)}{\sqrt{1+e^{2c}}} \quad . 
\]
Substituting into (\ref{4.5}b,c) we obtain the following set of differential equations for $a(z)$ and $b(z)$:
\begin{equation}\label{4.6}
\d a(z) /d z =  \frac{\mu\,e^c}{\sqrt{1+e^{2c}}} b(z) \quad ; \quad \d b(z) /d z = - \frac{\mu\,e^c}{\sqrt{1+e^{2c}}} a(z)\quad . 
\end{equation}
These equations are independent of the choice of $\alpha(z)$. Therefore, eqs.\eqref{4.6} do not put any restriction on the quantities $\alpha(z)$ (that are related to the vector field $X$), but result into a set of compatibility conditions on the boundary data. In fact, solving eqs.~\eqref{4.6} we obtain that \eqref{4.4} satisfies the field equations only if
\[
a(z) = \left[A\cos\left(\frac{\mu e^{2c}}{\sqrt{1+e^{2c}}}z\right) + B\sin\left(\frac{\mu e^{2c}}{\sqrt{1+e^{2c}}}z\right)\right] \quad , 
\]
\[
b(z) = \left[B\cos\left(\frac{\mu e^{2c}}{\sqrt{1+e^{2c}}}z\right) - A\sin\left(\frac{\mu e^{2c}}{\sqrt{1+e^{2c}}}z\right)\right] \quad . 
\]
where $A$ and $B$ are arbitrary constants. This means that, evaluating for $\xi=0$, the only possible boundary data are of the form
\[
f(z) = a(z) \quad ; \quad g(z) = \mu\, b(z) \quad . 
\]

\noindent
Substituting into eqs.~\eqref{4.4} we obtain the most general solutions of the field equations with the given initial data, which are derivable from the Hamilton--Jacobi approach. Even in this simple case, they are by no way the most general ones. 

\noindent
Integrating eqs.~(\ref{3.10}c) we obtain that $u_\mu\/(z,\xi) = \gamma_\mu = const.$; then the solution of the Hamilton--Jacobi equation can be obtained evaluating the integral
\[
u\/(z,\xi) = \gamma_\mu A^\mu\/(z)\, \xi + \alpha^2\/(z) \int \left(u_1\/(\xi,z)\right)^2 d\xi \quad . 
\]
After the rather tedious process of integration, the solution can be obtained taking eqs.~\eqref{4.2} into account and evaluating $x^1,x^2$ as functions of $z,\xi$ for any possible choice of the function $\alpha(z)$. We obtain that:
\[
S^1(x^1,x^2) = u\/(z(x^1,x^2),\xi(x^1,x^2))\,\frac{\alpha(z\/(x^1,x^2))}{\sqrt{1+e^{2c}}}\quad , 
\]
\[
S^2(x^1,x^2) = u\/(z(x^1,x^2),\xi(x^1,x^2))\,\frac{\alpha(z\/(x^1,x^2))\,e^c}{\sqrt{1+e^{2c}}}\quad . 
\]

\bibliographystyle{aip}

\end{document}